\UseRawInputEncoding
\documentclass[aps, prc, floatfix, nofootinbib, superscriptaddress, twocolumn]{revtex4-1}

\usepackage{latexsym}
\usepackage{amsmath}
\usepackage{amssymb}
\usepackage{amsfonts}

\usepackage[mathscr,scaled=1.15]{urwchancal}
\DeclareFontFamily{OT1}{pzc}{}
\DeclareFontShape{OT1}{pzc}{m}{it}%
{<-> s * [1.15] pzcmi7t}{}
\DeclareMathAlphabet{\mathpzc}{OT1}{pzc}{m}{it}

\usepackage{color}

\usepackage{supertabular}
\usepackage{placeins}
\usepackage{epsfig}
\usepackage{graphicx}
\usepackage{multirow}

\definecolor{purple}{rgb}{0.5,0,0.5}
\definecolor{blue}{rgb}{0.0,0,0.9}
\definecolor{prdblue}{rgb}{0.133,0.118,0.498}
\usepackage[colorlinks=true, pdfstartview=FitV, linkcolor=prdblue, citecolor= prdblue, urlcolor=prdblue]{hyperref}

\begin{document}

\title{Study of the $c\bar{c}s\bar{s}$ system in the chiral quark model}

\author{Xiaoyun Chen}
\email[]{xychen@jit.edu.cn}
\affiliation{College of Science, Jinling Institute of Technology, Nanjing 211169, P. R. China}

\author{Yue Tan}
\email[]{tanyue@ycit.edu.cn}
\affiliation{Department of Physics, Yancheng Institute of  Technology, Yancheng 224000, P. R. China}

\author{Xuejie Liu}
\email[]{1830592517@qq.com}
\affiliation{School of Physics, Henan Normal University, Xinxiang 453007, P. R. China}

\author{Jialun Ping}
\email[]{jlping@njnu.edu.cn}
\affiliation{College of Physics and Technology, Nanjing Normal University, Nanjing 211169, P. R. China}

\begin{abstract}
Recently, a charmonium $X(3960)$ in $B$ decays in the $D_s^+D_s^-$ invariant-mass spectrum is discovered by the LHCb Collaboration with the quantum number $J^{PC}=0^{++}$. Motivated by the discovery, in this work, we systematically investigated the $c\bar{c}s\bar{s}$ tetraquark states with the quantum numbers $J^{PC}=0^{++}, 1^{++}, 1^{+-}, 2^{++}$ in the framework of the chiral quark model(CQM). In our calculations, we considered the meson-meson structure of the tetraquark states and the diquark-antidiquark structure, as well as the channel-coupling of all channels of these two configurations are considered in this work. For example, all color structures including color singlet, hidden color channel, and the mixing of them are also taken into account. The numerical results indicates that no bound states were found in our model. But there exist several resonant states by using the stabilization method, the real scaling method (RSM) so called. Among these states, the $0^{++}$ resonant state with mass 3927 MeV matches very well with the energy of the newly discovered exotic state $X(3960)$ reported by the LHCb collaboration. As a result, our calculations suggest that $X(3960)$ can be interpreted as a $c\bar{c}s\bar{s}$ tetraquark state with quantum number $J^{PC}=0^{++}$. Apart form that, we also find several resonance states with mass 4179 MeV, 4376 MeV with $0^{++}$. For $1^{++}$, there is likely one resonance state in the energy range of 4310$\sim$4336 MeV, along with two resonance states at the energy of 4395 MeV and 4687 MeV, respectively. Besides, two resonance states at 4300 MeV and 4355 MeV for $1^{+-}$, as well as one state at 4788 MeV for 
$2^{++}$, are found, which are likely to be new exotic states. More experimental data is needed to confirm the existence of these resonance states.
\end{abstract}
\maketitle


\section{Introduction}
\label{introduction}
Quantum Chromodynamics (QCD), the fundamental theory describing the strong interaction, is non-perturbative at low energies. This makes it extremely challenging to solve problems using the hadron spectrum model alone. The traditional quark model effectively accounts for the hadron spectrum by classifying hadrons into mesons (composed of a quark and an antiquark, $q\bar{q}$) and baryons (composed of three quarks, $qqq$) \cite{GEllmann, Zweig}. However, numerous states and resonant structures observed in experiments over the past two decades do not align with the hadron spectrum predicted by the naive quark model. These states, known as exotic states, have prompted extensive efforts to understand their true nature, but their characteristics remain a subject of ongoing debate.

Recently, the LHCb Collaboration announced the observation of a new resonant structure $X(3960)$ in the $D_s^+D_s^-$ invariant mass distribution of the $B^+ \rightarrow D_s^+D_s^-K^+$ decay \cite{LHCb:2022aki}. The peak structure is just above $D_s^+D_s^-$ threshold with a statistical significance larger than $12\sigma$. The mass, width and the quantum numbers of the structure are measured to be $3935\pm5\pm10$ MeV, $43\pm13\pm8$ MeV and $J^{PC}=0^{++}$, respectively. The properties of the new structure are consistent with recent theoretical predictions for a state composed of $c\bar{c}s\bar{s}$ quarks. Evidence for an additional structure is found around 4140 MeV in the $D_s^+D_s^-$ invariant mass by LHCb, which might be caused either by a new resonance with the $0^{++}$ assignment or by a $J/\psi\phi \leftrightarrow D_s^+D_s^-$ coupled-channel effect.

Usually, if a new state is discovered in an invariant mass spectrum in an experiment, where the mass spectrum contains a pair of heavy and anti-heavy mesons, we usually assume that it is a conventional charmonium state as our first thought. Since the quantum number of $X(3960)$ is reported as $0^{++}$, the first opinion is whether it is a new $\chi_{c_0}$ state. So we check the theoretical masses of $\chi_{c_0}(1P),\chi_{c_0}(2P)$ and $\chi_{c_0}(3P)$ in the conventional quark model. Our results indicates $M(\chi_{c_0}(1P))=3362.8$ MeV, $M(\chi_{c_0}(2P))=3814.7$ MeV, $M(\chi_{c_0}(3P))=4290.9$ MeV, and the theoretical masses don't agree with the experiment of $X(3960)$. Besides, For $X(3915)$ state, although its mass is close to the $X(3960)$,due to its mass is below the threshold of $D_s\bar{D}_s$, about 3938 MeV, it is puzzled that it can be observed in the $D_s\bar{D}_s$ invariant mass spectrum.
Also, from the point of view of decay property, the LHCb Collaboration compared its decay widths to $D^+D^-$ and $D_s^+D_s^-$. The result gives that \cite{LHCb:2022aki}
\begin{eqnarray*}
\frac{\Gamma(X(3960) \rightarrow D^+D^-)}{\Gamma(X(3960) \rightarrow D_s^+D_s^-)}=0.29\pm0.09\pm0.10\pm0.008,
\end{eqnarray*}
which indicates that it is easier for $X(3960)$ to decay into $D_s^+D_s^-$ rather than $D^+D^-$. As we know, it's harder to create a strange quark pair $s\bar{s}$ in a vacuum compared with a light quark pair $u\bar{u}$ or $d\bar{d}$, so conventional charmonia predominantly decay into a pair of $D$ meson, which means $X(3960)$ may not a conventional charmonia state, but a new exotic state with quark composition $c\bar{c}s\bar{s}$, since a state composed by four valence quarks may be the easiest generalization.

Since the mass of $X(3960)$ is close to the $D_s\bar{D}_s$ threshold, a consideration that it is related to some molecular states can easily raise up \cite{Xin:2022bzt,Ji:2022uie}.  In addition, Ref. \cite{Ji:2022uie} pointed out that virtual state explanation and the bound state interpretation, were also valid. Then, Ref. \cite{Xie:2022lyw} used the effective Lagrangian approach to calculate the production rate of $X(3960)$ in the B decays utilizing triangle diagrams, and the results indicates that both the virtual and bound state interpretations can match the relevant experimental data. In addition, there are many theoretical explanations for $X(3960)$. For example, some references such as \cite{Prelovsek:2020eiw, Gamermann:2006nm, Nieves:2012tt, Hidalgo-Duque:2012rqv, Meng:2020cbk, Dong:2021juy, Bayar:2022dqa, Xin:2022bzt, Ji:2022uie,Xie:2022lyw} explain the state as an effect caused by a molecular state located below the $D_s^+D_s^-$ threshold, since $X(3960)$ is observed in the $D_s^+D_s^-$ invariant mass spectrum. However, given that its measurement mass is higher than the $D_s^+D_s^-$ threshold \cite{LHCb:2022aki}, we believe that the explanation of its resonance state is also possible.

Stimulated by the results of $X(3960)$ by recent LHCb experiments, which may a tetraquark state $c\bar{c}s\bar{s}$, we have a great interest in studying the tetraquark system composed of $c\bar{c}s\bar{s}$. In fact, many related resonances have been found experimentally, such as $X(4140)$ \cite{CDF:2009jgo, LHCb:2012wyi, CMS:2013jru, D0:2013jvp, BaBar:2014wwp}, $X(4350)$ \cite{Belle:2009rkh}, $X(4274)$ \cite{CDF:2011pep}, $X(4140)$ and $X(4274)$ \cite{LHCb:2016axx}, $X(4500)$ and $X(4700)$ \cite{LHCb:2016nsl}, $X(4685)$ and $X(4630)$ \cite{LHCb:2021uow}. Theoretically, a lot of work has been done on these states. For example, as early as 2009, Stancu has calculated the spectrum of tetraquark of type $c\bar{c}s\bar{s}$ within a simple quark model with chromomagnetic interaction and effective quark masses extracted from meson and baryon spectra. The mass of the lowest $0^{++}$ state was found to be 3995 MeV \cite{Stancu:2009ka}. Ref. \cite{Ortega:2016hde} thought the $X(4140)$ resonance as a cusp and the $X(4274)$, $X(4500)$ and $X(4700)$ were all regarded as the conventional charmonium $c\bar{c}$ states with a nonrelativistic constituent quark model. While in the relativized quark model \cite{Lu:2016cwr}, the resonance of $X(4140)$ was defined as the $c\bar{c}s\bar{s}$ tetraquark ground state; and $X(4274)$ was a good candidate of the conventional $\chi_{c_1}$ state; $X(4500)$ and $X(4700)$ were explained as highly excited tetraquark states. Besides, in Ref. \cite{Yang:2019dxd}, $X(4274)$ was defined as the $c\bar{c}s\bar{s}$ tetraquark state with $J^{PC}=1^{++}$, $X(4350)$ as a good candidate of the compact tetraquark sate with $J^{PC}=0^{++}$, the $X(4700)$ as the $2S$ radial excited tetraquark state with $J^{PC}=0^{++}$. In the framework of the multiquark color flux-tube model \cite{Deng:2017xlb}, the authors gave the results that the $X(4500)$ and the $X(4700)$ were $S-$wave radial excited states $[cs][\bar{c}\bar{s}]$. Based on the diquark-antidiquark configuration in the QCD sum rules \cite{Chen:2016oma}, the $X(4500)$ and the $X(4700)$ were indicated as the $D-$wave $cs\bar{c}\bar{s}$ tetraquark states of $J^P=0^+$.

In this work, we systematically study the properties of the $c\bar{c}s\bar{s}$ system with quantum numbers $J^{PC}=0^{++}, 1^{++}, 1^{+-}, 2^{++}$ by using the chiral quark model(CQM). In the present calculation, two configurations, the meson-meson ($q\bar{q}-q\bar{q}$) and the diquark-antidiquark ($qq-\bar{q}\bar{q}$), are taken into account. Besides, to be more convincing, the channel coupling effect of the $c\bar{c}s\bar{s}$ tetraquark systems is also included.

This work is organized as follows. In Sect. \ref{ModleandGEM}, we present a review of the chiral quark model and the wave functions of the total system in the present work. The numerical results and a discussion for the tetraquarks are given in Sect. \ref{numerical}. Finally, the last section is devoted to a brief summary.


\section{Theoretical framework}
\label{ModleandGEM}
\subsection{The chiral quark model(CQM)}
Since 2003, various theoretical approaches have been employed to explore the properties of multiquark candidates observed in experiments. Among these, the QCD-inspired quark model remains a powerful and straightforward tool for describing hadron spectra and hadron-hadron interactions, achieving significant success. This model has been applied in our previous studies to examine tetraquark systems, yielding valuable insights \cite{Chen:2016npt,Chen:2018hts,Chen:2019vrj}. In this context, the chiral quark model holds great promise for investigating doubly heavy tetraquark states, such as $c\bar{c}s\bar{s}$.

The Hamiltonian of the chiral quark model can be written as follows for four-body system,
\begin{align}
 H & = \sum_{i=1}^4 m_i  +\frac{p_{12}^2}{2\mu_{12}}+\frac{p_{34}^2}{2\mu_{34}}
  +\frac{p_{1234}^2}{2\mu_{1234}}  \quad  \nonumber \\
  & + \sum_{i<j=1}^4 \left[ V_{ij}^{C}+V_{ij}^{G}+\sum_{\chi=\pi,K,\eta} V_{ij}^{\chi}
   +V_{ij}^{\sigma}\right].
\end{align}
The potential energy: $V_{ij}^{C, G, \chi, \sigma}$ represents the confinement, one-gluon-exchange(OGE), Goldston boson exchange and scalar $\sigma$ meson-exchange,
respectively. According to Casimir scheme, the forms of these potentials can be directly extended to multiqaurk systems with the Casimir factor
$\boldsymbol{\lambda}_i \cdot \boldsymbol{\lambda}_j $~\cite{Casimir}. Their forms are:
{\allowdisplaybreaks
\begin{subequations}
\begin{align}
V_{ij}^{C}&= ( -a_c r_{ij}^2-\Delta ) \boldsymbol{\lambda}_i^c
\cdot \boldsymbol{\lambda}_j^c ,  \\
 V_{ij}^{G}&= \frac{\alpha_s}{4} \boldsymbol{\lambda}_i^c \cdot \boldsymbol{\lambda}_{j}^c
\left[\frac{1}{r_{ij}}-\frac{2\pi}{3m_im_j}\boldsymbol{\sigma}_i\cdot
\boldsymbol{\sigma}_j
  \delta(\boldsymbol{r}_{ij})\right],  \\
\delta{(\boldsymbol{r}_{ij})} & =  \frac{e^{-r_{ij}/r_0(\mu_{ij})}}{4\pi r_{ij}r_0^2(\mu_{ij})}, \\
V_{ij}^{\pi}&= \frac{g_{ch}^2}{4\pi}\frac{m_{\pi}^2}{12m_im_j}
  \frac{\Lambda_{\pi}^2}{\Lambda_{\pi}^2-m_{\pi}^2}m_\pi v_{ij}^{\pi}
  \sum_{a=1}^3 \lambda_i^a \lambda_j^a,  \\
V_{ij}^{K}&= \frac{g_{ch}^2}{4\pi}\frac{m_{K}^2}{12m_im_j}
  \frac{\Lambda_K^2}{\Lambda_K^2-m_{K}^2}m_K v_{ij}^{K}
  \sum_{a=4}^7 \lambda_i^a \lambda_j^a,   \\
\nonumber
V_{ij}^{\eta} & =
\frac{g_{ch}^2}{4\pi}\frac{m_{\eta}^2}{12m_im_j}
\frac{\Lambda_{\eta}^2}{\Lambda_{\eta}^2-m_{\eta}^2}m_{\eta}
v_{ij}^{\eta}  \\
 & \quad \times \left[\lambda_i^8 \lambda_j^8 \cos\theta_P
 - \lambda_i^0 \lambda_j^0 \sin \theta_P \right],   \\
 v_{ij}^{\chi} & =  \left[ Y(m_\chi r_{ij})-
\frac{\Lambda_{\chi}^3}{m_{\chi}^3}Y(\Lambda_{\chi} r_{ij})
\right]
\boldsymbol{\sigma}_i \cdot\boldsymbol{\sigma}_j,\\
V_{ij}^{\sigma}&= -\frac{g_{ch}^2}{4\pi}
\frac{\Lambda_{\sigma}^2}{\Lambda_{\sigma}^2-m_{\sigma}^2}m_\sigma \nonumber \\
& \quad \times \left[
 Y(m_\sigma r_{ij})-\frac{\Lambda_{\sigma}}{m_\sigma}Y(\Lambda_{\sigma} r_{ij})\right]  ,
\end{align}
\end{subequations}}
\hspace*{-0.5\parindent}%
where $Y(x)  =   e^{-x}/x$ is the standard Yukawa function;
$\{m_i\}$ are the constituent masses of quarks and antiquarks, and $\mu_{ij}$ are their reduced masses;
\begin{equation}
\mu_{1234}=\frac{(m_1+m_2)(m_3+m_4)}{m_1+m_2+m_3+m_4};
\end{equation}
$\mathbf{p}_{ij}=(\mathbf{p}_i-\mathbf{p}_j)/2$, $\mathbf{p}_{1234}= (\mathbf{p}_{12}-\mathbf{p}_{34})/2$;
$r_0(\mu_{ij}) =s_0/\mu_{ij}$;
$\boldsymbol{\sigma}$ are the $SU(2)$ Pauli matrices;
$\boldsymbol{\lambda}$, $\boldsymbol{\lambda}^c$ are $SU(3)$ flavor, color Gell-Mann matrices, respectively;
$g^2_{ch}/4\pi$ is the chiral coupling constant, determined from the $\pi$-nucleon coupling;
and $\alpha_s$ is an effective scale-dependent running coupling \cite{Valcarce:2005em},
\begin{equation}
\alpha_s(\mu_{ij})=\frac{\alpha_0}{\ln\left[(\mu_{ij}^2+\mu_0^2)/\Lambda_0^2\right]}.
\end{equation}
It is worth noting that, in this work, we focus on the low-lying positive parity $c\bar{c}s\bar{s}$ tetraquark states of $S-$wave, and the spin-orbit and tensor interactions are not included. The interactions involving Goldstone-boson exchange between light quarks arise as a consequence of the dynamical breaking of chiral symmetry. In the $c\bar{c}s\bar{s}$ system, the $\pi$ and $K$ exchange interactions are absent due to the lack of up or down quarks. Instead, only the $\eta$ exchange term is effective between the $s\bar{s}$ pair.

In this work, the model parameters are obtained by fitting the meson spectra across a range from light to heavy, with the resulting values presented in Table~\ref{modelparameters}. With these model parameters, we get the corresponding meson spectra of $J/\psi$, $\phi$, $\eta_c$, $\eta^{\prime}$, $D_s^{(*)}$ and $\bar{D}_s^{(*)}$, which are list in Table ~\ref{mesonspectra}. In comparison with experiments, we can see that the quark model can successfully describe the hadron spectra. Then we use these model parameters to investigate the double heavy $c\bar{c}s\bar{s}$ systems.

\linespread{1.2}
\begin{table}[!t]
\begin{center}
\caption{ \label{modelparameters}
Model parameters, determined by fitting the meson spectra.}
\begin{tabular}{llr}
\hline\hline\noalign{\smallskip}
Quark masses   &$m_u=m_d$    &313  \\
   (MeV)       &$m_s$         &536  \\
               &$m_c$         &1728 \\
               &$m_b$         &5112 \\
\hline
Goldstone bosons   &$m_{\pi}$     &0.70  \\
   (fm$^{-1} \sim 200\,$MeV )     &$m_{\sigma}$     &3.42  \\
                   &$m_{\eta}$     &2.77  \\
                   &$m_{K}$     &2.51  \\
                   &$\Lambda_{\pi}=\Lambda_{\sigma}$     &4.2  \\
                   &$\Lambda_{\eta}=\Lambda_{K}$     &5.2  \\
                   \cline{2-3}
                   &$g_{ch}^2/(4\pi)$                &0.54  \\
                   &$\theta_p(^\circ)$                &-15 \\
\hline
Confinement        &$a_c$ (MeV fm$^{-2}$)         &101 \\
                   &$\Delta$ (MeV)     &-78.3 \\
\hline
OGE                 & $\alpha_0$        &3.67 \\
                   &$\Lambda_0({\rm fm}^{-1})$ &0.033 \\
                  &$\mu_0$(MeV)    &36.98 \\
                   &$s_0$(MeV)    &28.17 \\
\hline\hline
\end{tabular}
\end{center}
\end{table}

\linespread{1.2}
\begin{table}[!t]
\begin{center}
\renewcommand\tabcolsep{6.0pt} 
\caption{ \label{mesonspectra} The mass spectra of $c\bar{c}$, $s\bar{s}$, $c\bar{s}$ in the chiral quark model in comparison with the experimental data \cite{PDG} (in unit of MeV).}
\begin{tabular}{ccccc}
\hline\hline\noalign{\smallskip}
State       &$I(J^P)$  &Energy &Meson &Expt \cite{PDG}\\
\hline
$c\bar{c}$      &$0(1^-)$  &1S: 3096.7   &$J/\psi$  &3096.9 \\
                                          &                    &2S: 3605.2  &...    &...       \\
                                          &                    &3S: 4202.8   &...  &...       \\
            &$0(0^-)$      &1S: 2964.5   &$\eta_c$   &2983.6\\
                                          &                    &2S: 3509.0   &...   &...       \\
                                         &                    &3S: 4062.9    &... &...       \\
$s\bar{s}$  &$0(1^-)$  &1S: 1016.1     &$\phi$ &1019.4 \\
                               &                    &2S: 1889.6     &...   &...    \\
                               &                    &3S: 2762.3    &...    &...   \\
                               &                    &4S: 4387.7    &...    &...   \\
                         &$0(0^-)$  &1S: 821.5    &$\eta^{\prime}$     &957.8\\
                         &                    &2S:1712.9     &...   &...    \\
                         &                    &3S:2541.8    &...    &...   \\
                         &                    &4S:4229.6    &...    &...   \\
$c\bar{s}$&$0(0^-)$  &1S: 1950.1    &$D_s$   &1968.3 \\
                            &                    &2S: 2664.6     &...   &...    \\
                         &                    &3S: 3337.5    &...    &...   \\
                         &                    &4S: 4121.9    &...    &...   \\
            &$0(1^-)$  &1S: 2079.9    &$\bar{D}_s^*$     &2112.2\\
                        &                    &2S: 2778.9     &...   &...    \\
                         &                    &3S: 3455.9    &...    &...   \\
                         &                    &4S: 4229.8    &...    &...   \\
\hline\hline
\end{tabular}
\end{center}
\end{table}

\subsection{The wave function}
\begin{figure}
\center{\includegraphics{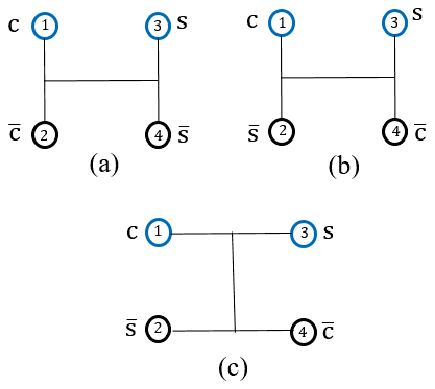}}
\caption{\label{configure} Two types of configurations in $c\bar{c}s\bar{s}$ tetraquarks. (a) and (b) is the meson-meson structure, and (c) is the diquark-antidiquark configuration. }
\end{figure}

To obtain the wave functions of the tetraquark $c\bar{c}s\bar{s}$ state, the resonance group method \cite{Kamimura:1977okl} is applied. For $c\bar{c}s\bar{s}$ system, two kinds of configurations are represented in Figure \ref{configure}, which are meson-meson structures shown in Fig. \ref{configure}(a) and \ref{configure}(b), and the diquark-antidiquark structure shown in Fig. \ref{configure}(c). To simplify the challenging four-body problem, the current calculation focuses solely on these two structures. However, an efficient approach is employed to combine these configurations and evaluate the impact of multi-channel coupling. At the quark level, four fundamental degrees of freedom: color, spin, flavor, and orbit, are widely recognized by QCD theory. The wave function of the multiquark system is constructed as a direct product of the color, spin, flavor, and orbit components, that contribute to a given well defined quantum numbers $I(J^{PC})$.

\subsubsection{the color wave function}
For the meson-meson picture, the indices of particles are ``1234'', and for the diquark-antidiquark picture, the indices are ``1324''.
For the color part, in the meson-meson picture, the colorless wave functions can be obtained from $\big[[c\bar{c}]_{1_c}[s\bar{s}]_{1_c}\big]_1$ or $\big[[c\bar{c}]_{8_c}[s\bar{s}]_{8_c}\big]_1$. In the diquark-antidiquark picture, the color representation of the diquark maybe antisymmetrical $[cs]_{\bar{3}_c}$ or symmetrical $[cs]_{6_c}$, and for antidiquark, the color form is antisymmetrical $[\bar{c}\bar{s}]_{3_c}$ or symmetrical $[\bar{c}\bar{s}]_{\bar{6}_c}$. There are two rules to couple the diquark and antidiquark into a colorless wave function: one is the good diquark with attractive interaction $\big[[cs]_{\bar{3}_c}[\bar{c}\bar{s}]_{3_c}\big]_1$, and another is the bad diquark with repulsive interaction $\big[[cs]_{6_c}[\bar{c}\bar{s}]_{\bar{6}_c}\big]_1$.
So we can easily write down the color wave functions in the meson-meson picture and the diquark-antidiquark picture, respectively.
\begin{subequations}
\begin{align}
\chi^{c_1}_{1 \otimes 1}&=\frac{1}{3}(r\bar{r}+g\bar{g}+b\bar{b})_{12}(r\bar{r}+g\bar{g}+b\bar{b})_{34}, \\
\chi^{c_2}_{8 \otimes 8}&=\frac{\sqrt{2}}{12}(3r\bar{b}b\bar{r}+3r\bar{g}g\bar{r}+3g\bar{b}b\bar{g}+3b\bar{g}g\bar{b} \nonumber \\
&+3g\bar{r}r\bar{g}+3b\bar{r}r\bar{b}+2r\bar{r}r\bar{r}+2g\bar{g}g\bar{g} \nonumber \\
&+2b\bar{b}b\bar{b}-r\bar{r}g\bar{g}-g\bar{g}r\bar{r}-b\bar{b}g\bar{g} \nonumber \\
&-b\bar{b}r\bar{r}-g\bar{g}b\bar{b}-r\bar{r}b\bar{b})_{1234}, \\
\chi^{c_3}_{\bar{3} \otimes 3}&=\frac{\sqrt{3}}{6}(rg\bar{r}\bar{g}-rg\bar{g}\bar{r}+gr\bar{g}\bar{r}-gr\bar{r}\bar{g} \nonumber \\
&+rb\bar{r}\bar{b}-rb\bar{b}\bar{r}+br\bar{b}\bar{r}-br\bar{r}\bar{b} \nonumber \\
&+gb\bar{g}\bar{b}-gb\bar{b}\bar{g}+bg\bar{b}\bar{g}-bg\bar{g}\bar{b})_{1324},  \\
\chi^{c_4}_{6 \otimes \bar{6}}&=\frac{\sqrt{6}}{12}(2rr\bar{r}\bar{r}+2gg\bar{g}\bar{g}+2bb\bar{b}\bar{b}+rg\bar{r}\bar{g}\nonumber \\
&+rg\bar{g}\bar{r}+gr\bar{g}\bar{r}+gr\bar{r}\bar{g}+rb\bar{r}\bar{b}\nonumber \\
&+rb\bar{b}\bar{r}+br\bar{b}\bar{r}+br\bar{r}\bar{b}+gb\bar{g}\bar{b}\nonumber \\
&+gb\bar{b}\bar{g}+bg\bar{b}\bar{g}+bg\bar{g}\bar{b})_{1324}.
\end{align}
\end{subequations}

\subsubsection{the spin wave function}
For the spin, the total spin $S$ of the tetraquark states can vary between 0 and 2, and all possible values are taken into account. The wave functions for the two-body clusters are given as follows:
\begin{subequations}
\begin{align}
\chi_{11}&=\alpha\alpha,\\
\chi_{10}&=\frac{1}{\sqrt{2}}(\alpha\beta+\beta\alpha),\\
\chi_{1-1}&=\beta\beta, \\
\chi_{00}&=\frac{1}{\sqrt{2}}(\alpha\beta-\beta\alpha).
\end{align}
\end{subequations}
If the spin of one cluster is coupled  to $S_1$ and that of another cluster to $S_2$, the total spin wave function of the four-quark system can be obtained as $S=S_1 \oplus S_2$,
\begin{subequations}
\begin{align}
\chi_{00=0\oplus0}^{\sigma_1}&=\chi_{00}\chi_{00}, \\
\chi_{00=1\oplus1}^{\sigma_2}&=\sqrt{\frac{1}{3}}(\chi_{11}\chi_{1-1}-\chi_{10}\chi_{10}+\chi_{1-1}\chi_{11}), \\
\chi_{11=0\oplus1}^{\sigma_3}&=\chi_{00}\chi_{11}, \\
\chi_{11=1\oplus0}^{\sigma_4}&=\chi_{11}\chi_{00},\\
\chi_{11=1\oplus1}^{\sigma_5}&=\frac{1}{\sqrt{2}}(\chi_{11}\chi_{10}-\chi_{10}\chi_{11}),\\
\chi_{22=1\oplus1}^{\sigma_6}&=\chi_{11}\chi_{11}.
\end{align}
\end{subequations}
The subscript of $\chi$ represents the  $SM_S=S_1 \oplus S_2$, and $M_S$ is the third projection of the total spin $S$.

\subsubsection{the flavor wave function}
Regarding the flavor degree of freedom, the tetraquark systems consist of two heavy quarks and two strange quarks, placing them in the isoscalar sector with $I=0$. The flavor wave functions, represented as $\chi_{I}^{f_i}$, with the subscript $I$ referring to isoscalar, can be expressed as follows:
\begin{subequations}
\begin{align}
\chi_{0}^{f_1}&=c\bar{c}s\bar{s}, \\
\chi_{0}^{f_2}&=c\bar{s}s\bar{c}, \\
\chi_{0}^{f3}&=cs\bar{c}\bar{s}.
\end{align}
\end{subequations}

\subsubsection{the orbital wave function}
For orbital part, in our calculations, the orbital wave function is
\begin{equation}\label{spatialwavefunctions}
\Psi_{L}^{M_{L}}=\left[[\Psi_{l_1}({\bf r})\Psi_{l_2}({\bf
R})]_{l_{12}}\Psi_{L_r}(\bf{Z}) \right]_{L}^{M_{L}},
\end{equation}
where, $\bf{r}$, $\bf{R}$ and $\bf{Z}$ are the relative spatial coordinates, and one of the definitions of the Jacobi coordinates can be written as,
\begin{align}
\bf{r}&=r_1-r_2, \nonumber \\
\bf{R}&=r_3-r_4, \nonumber \\
\bf{Z}&=\frac{m_1{\bf r}_1+m_2{\bf r}_2}{m_1+m_2}-\frac{m_3{\bf r}_3+m_4{\bf r}_4}{m_3+m_4}, \nonumber \\
\bf{R}_c&=\frac{m_1{\bf r}_1+m_2{\bf r}_2+m_3{\bf r}_3+m_4{\bf r}_4}{m_1+m_2+m_3+m_4}.
\end{align}
$\bf{R}_c$ is the center-of-mass coordinate. In Eq. (\ref{spatialwavefunctions}), $l_1$, $l_2$ is the inner angular momentum of the two sub-cluster; $L_r$ is the relative angular momentum between two sub clusters. $L$ is the total orbital angular momentum of the four-quark system, with $L=l_1 \oplus l_2 \oplus L_r$. In the present work, we just consider the low-lying $S\mbox{-}$wave double heavy tetraquark states, so it is natural to assume that all the orbital angular momenta are zeros. The parity of the double-heavy tetraquarks $c\bar{c}s\bar{s}$ can be expressed in terms of the relative orbital angular momenta, with $P=(-1)^{l_1+l_2+L_r}=+1$. So as to get the reliable information of the four-quark system, a high precision numerical method, Gaussian expansion method~(GEM) \cite{Hiyama:2003cu} is applied in our work. In GEM, any relative motion wave function can be expanded in series of Gaussian basis functions,
\label{radialpart}
\begin{align}
\Psi_{l}^{m}(\mathbf{x}) & = \sum_{n=1}^{n_{\rm max}} c_{n}N_{nl}x^{l}
e^{-\nu_{n}x^2}Y_{lm}(\hat{\mathbf{x}}),
\end{align}
where $N_{nl}$ are normalization constants,
\begin{align}
N_{nl}=\left[\frac{2^{l+2}(2\nu_{n})^{l+\frac{3}{2}}}{\sqrt{\pi}(2l+1)}
\right]^\frac{1}{2}.
\end{align}
$c_n$ are the variational parameters, which are determined dynamically. The Gaussian size
parameters are chosen according to the following geometric progression
\begin{equation}\label{gaussiansize}
\nu_{n}=\frac{1}{r^2_n}, \quad r_n=r_1a^{n-1}, \quad
a=\left(\frac{r_{n_{\rm max}}}{r_1}\right)^{\frac{1}{n_{\rm
max}-1}}.
\end{equation}
This procedure enables optimization of the expansion using just small numbers of Gaussians. For example, in order to obtain the stable ground-state masses of mesons in Table \ref{mesonspectra}, we takes,
\begin{equation}
r_1=0.01~\rm{fm}, \quad \emph{r}_{\emph{n}_{\rm max}}=2~\rm{fm}, \quad \emph{n}_{\rm{max}}=12.
\end{equation}

Finally, to fulfill the Pauli principle, the complete wave function is written as
\begin{equation}\label{totalwavefunction}
\Psi=\mathscr{A}[\Psi_{L}^{M_{L}}\chi_{SM_S}^{\sigma_j}]_{JM_{J}}\chi_{0}^{f_i}\chi^{c_k},
\end{equation}
where $\mathscr{A}$ is the antisymmetry operator of double-heavy tetraquarks. If the two quarks or two antiquarks in the tetraquark are identical particles, $\mathscr{A} = \frac{1}{2}(1-P_{13}-P_{24}+P_{13}P_{24})$. In this work, the operator $\mathscr{A}$ is defined as $\mathscr{A} = 1$ due to the absence of any identical quarks in the $c\bar{c}s\bar{s}$ system.

\section{numerical analysis}
\label{numerical}
In this study, the low-lying S-wave states of the $c\bar{c}s\bar{s}$ tetraquark are examined systematically. Assuming that the total orbital angular momentum $L$ is 0, the parity of the $c\bar{c}s\bar{s}$ tetraquark is positive. Consequently, the total angular momentum $J$ can take values of 0, 1, or 2. The isospin value for the $c\bar{c}s\bar{s}$ tetraquark system is constrained to be 0. Two possible structures for the $c\bar{c}s\bar{s}$ tetraquark, namely meson-meson and diquark-antidiquark, are investigated. In each structure, all possible states are considered, which are list in Table \ref{channels} For meson-meson structure, two color configurations, color singlet-singlet($1\times1$) and octet-octet($8\times8$) are employed. For diquark-antidiquark structure, also two color configurations, color antitriplet-triplet($\bar{3}\times3$) and sexet-antisexet($6\times\bar{6}$) are taken into account.

\begin{table*}[!t]
\begin{center}
\renewcommand\tabcolsep{5.6pt} %
\caption{ \label{channels} All possible channels for all quantum numbers. In the table, $[i,j,k]$ represents the flavor, spin and color channels $[\chi_{0}^{f_{i}}, \chi_{SM_S}^{\sigma_{j}}, \chi^{c_{k}}]$.}
\begin{tabular}{ccccccccccccccc}
\hline\hline\noalign{\smallskip} 
\multicolumn{3}{c}{$J^{PC}=0^{++}$}   & &\multicolumn{3}{c}{$J^{PC}=1^{++}$} & &\multicolumn{3}{c}{$J^{PC}=1^{+-}$}  &     &\multicolumn{3}{c}{$J^{PC}=2^{++}$}   \\
\cline{1-3} \cline{5-7} \cline{9-11} \cline{13-15}\\
Index  &$[i,j,k]$ &channels &
&Index &$[i,j,k]$ &channels & 
&Index &$[i,j,k]$ &channels & 
&Index &$[i,j,k]$ &channels \\
\hline
1  &[1,1,1] &$\eta_c\eta^{\prime}$ &   &1  &[1,5,1] &$J/\psi\phi$         &&1  &[1,3,1] &$\eta_c\phi$          &  &1  &[1,6,1]  &$J/\psi\phi$       \\
2  &[1,2,1] &$J/\psi\phi$          &   &2  &[1,5,2] &...                  &&2  &[1,4,1] &$J/\psi\eta^{\prime}$ &  &2  &[1,6,2]  &...                \\
3  &[1,1,2] &...                   &   &3  &[2,5,1] &$D_s^{*+}D_s^{*-}$   &&3  &[1,3,2] &...                   &  &3  &[2,6,1]  &$D_s^{*+}D_s^{*-}$  \\
4  &[1,2,2] &...                   &   &4  &[2,5,2] &...                  &&4  &[1,4,2] &...                   &  &4  &[2,6,2]  &...                \\
5  &[2,1,1] &$D_s^+D_s^-$          &   &5  &[3,5,3] &$cs\bar{c}\bar{s}$   &&5  &[2,3,1] &$D_s^+D_s^{*-}$       &  &5  &[3,6,1]  &$cs\bar{c}\bar{s}$ \\
6  &[2,2,1] &$D_s^{*+}D_s^{*-}$    &   &6  &[3,5,4] &$cs\bar{c}\bar{s}$   &&6  &[2,4,1] &$D_s^{*+}D_s^-$       &  &6  &[3,6,2]  &$cs\bar{c}\bar{s}$ \\
7  &[2,1,2] &...                   &   &   &        &                     &&7  &[2,3,2] &...                   &  &   &         &                   \\
8  &[2,2,2] &...                   &   &   &        &                     &&8  &[2,4,2] &...                   &  &   &         &                   \\
9  &[3,1,3] &$cs\bar{c}\bar{s}$    &   &   &        &                     &&9  &[3,3,3] &$cs\bar{c}\bar{s}$    &  &   &         &                   \\
10 &[3,2,3] &$cs\bar{c}\bar{s}$    &   &   &        &                     &&10 &[3,4,3] &$cs\bar{c}\bar{s}$    &  &   &         &                   \\
11 &[3,1,4] &$cs\bar{c}\bar{s}$    &   &   &        &                     &&11 &[3,3,4] &$cs\bar{c}\bar{s}$    &  &   &         &                   \\
12 &[3,2,4] &$cs\bar{c}\bar{s}$    &   &   &        &                     &&12 &[3,4,4] &$cs\bar{c}\bar{s}$    &  &   &         &                   \\
\hline\hline
\end{tabular}
\end{center}
\end{table*}

The energy of the $c\bar{c}s\bar{s}$ system with quantum numbers  $J^{PC}=0^{++}, 1^{++}, 1^{+-}, 2^{++}$ for both the meson-meson and diquark-antidiquark structures, with the channel coupling of these two configurations are shown in Table \ref{0++}, \ref{1++}, \ref{1+-}, \ref{2++}, respectively. In the tables, the first column represents the index of each possible channel. The second and the third column is the physical channels. The fourth column stands for the theoretical threshold in the chiral quark model, which can be obtained by the values in the Table \ref{mesonspectra}. $E_{sc}$ is the energy for the every single channel. $E_{cc}$ represents the energy by considering the coupling of all channels for the meson-meson structure and the diquark-antidiquark structure, respectively. The last column $E_{mix}$ is the lowest energy of the $c\bar{c}s\bar{s}$ system by coupling all channels for these two structures. 

\begin{table}[!t]
\begin{center}
\caption{ \label{0++} The lowest-lying energies of $c\bar{c}s\bar{s}$ tetraquark system with $J^{PC}=0^{++}$ (in unit of MeV).}
\begin{tabular}{ccccccc}
\hline\hline\noalign{\smallskip} 
Index &[i;j;k] &channel             &Threshold  &\multicolumn{3}{c}{Energy} \\
      &        &      &                 &$E_{sc}$ &$E_{cc}$ &$E_{mix}$\\
\hline
1   &[1,1,1]&$\eta_c\eta^{\prime}$ &3785.9   &3791.6 &3788.5&3788.4\\
2   &[1,2,1]&$J/\psi\phi$          &4112.8   &4114.9 &  & \\
3   &[1,1,2]&...                   &         &4371.6 &  &\\
4   &[1,2,2]&...                   &         &4282.3 &  &\\
5   &[2,1,1]&$D_s^+D_s^-$          &3900.2   &3902.3       &  &\\
6   &[2,2,1]&$D_s^{*+}D_s^{*-}$    &4159.8   &4161.9       &  &\\
7   &[2,1,2]&...                   &         &4376.4       &  &\\
8   &[2,2,2]&...                   &         &4304.1       &  &\\
9   &[3,1,3]&$cs\bar{c}\bar{s}$    &         &4331.7 &4249.0&\\
10  &[3,2,3]&$cs\bar{c}\bar{s}$    &         &4384.3 &  &\\
11  &[3,1,4]&$cs\bar{c}\bar{s}$    &         &4386.1 &  &\\
12  &[3,2,4]&$cs\bar{c}\bar{s}$    &         &4305.3 &  &\\
\hline\hline
\end{tabular}
\end{center}
\end{table}

\begin{table}[!t]
\begin{center}
\caption{ \label{1++} The lowest-lying energies of $c\bar{c}s\bar{s}$ tetraquark system with $J^{PC}=1^{++}$ (in unit of MeV).}
\begin{tabular}{ccccccc}
\hline\hline\noalign{\smallskip} 
Index &[i;j;k] &channel             &Threshold  &\multicolumn{3}{c}{Energy} \\
      &      &  &                       &$E_{sc}$ &$E_{cc}$ &$E_{mix}$\\
\hline
1  &[1,5,1]&$J/\psi\phi$           &4112.8 &4115.3  &4115.3   &4115.3      \\
2  &[1,5,2]&...                    &       &4311.5  &&\\
3  &[2,5,1]&$D_s^{*+}D_s^{*-}$     &4159.8 &4161.9  &&\\
4  &[2,5,2]&...                    &       &4336.8  &&\\
5  &[3,5,3]&$cs\bar{c}\bar{s}$     &       &4390.5  &4326.4& \\
6  &[3,5,4]&$cs\bar{c}\bar{s}$     &       &4336.9  && \\
\hline\hline
\end{tabular}
\end{center}
\end{table}

\begin{table}[!t]
\begin{center}
\caption{ \label{1+-} The lowest-lying energies of $c\bar{c}s\bar{s}$ tetraquark system with $J^{PC}=1^{+-}$ (in unit of MeV).}
\begin{tabular}{ccccccc}
\hline\hline\noalign{\smallskip} 
Index &[i;j;k] &channel             &Threshold  &\multicolumn{3}{c}{Energy} \\
      &      &  &                       &$E_{sc}$ &$E_{cc}$ &$E_{mix}$\\
\hline
1  &[1,3,1]&$\eta_c\phi$           &3980.5 &3983.1&3920.7&3920.7\\
2  &[1,4,1]&$J/\psi\eta^{\prime}$  &3918.2 &3920.7  &&      \\
3  &[1,3,2]&...                    &       &4343.3  &&      \\
4  &[1,4,2]&...                    &       &4369.5  &&\\
5  &[2,3,1]&$D_s^+D_s^{*-}$        &4030.0 &4032.1  &&\\
6  &[2,4,1]&$D_s^{*+}D_s^-$        &4030.0 &4032.1  &&\\
7  &[2,3,2]&...                    &       &4371.8  &&\\
8  &[2,4,2]&...                    &       &4371.8  &&\\
9  &[3,3,3]&$cs\bar{c}\bar{s}$     &       &4362.9  &4323.5& \\
10 &[3,4,3]&$cs\bar{c}\bar{s}$     &       &4362.9  && \\
11 &[3,3,4]&$cs\bar{c}\bar{s}$     &       &4376.2  && \\
12 &[3,4,4]&$cs\bar{c}\bar{s}$     &       &4376.2  && \\
\hline\hline
\end{tabular}
\end{center}
\end{table}

\begin{table}[!t]
\begin{center}
\caption{ \label{2++} The lowest-lying energies of $c\bar{c}s\bar{s}$ tetraquark system with $J^{PC}=2^{++}$ (in unit of MeV).}
\begin{tabular}{ccccccc}
\hline\hline\noalign{\smallskip} 
Index &[i;j;k] &channel             &Threshold  &\multicolumn{3}{c}{Energy} \\
      &        &      &                 &$E_{sc}$ &$E_{cc}$ &$E_{mix}$\\
\hline
1  &[1,6,1]  &$J/\psi\phi$       &4112.8  &4114.9&4115.2&4115.2\\
2  &[1,6,2]  &...                &        &4367.6&  & \\
3  &[2,6,1]  &$D_s^{*+}D_s^{*-}$ &4159.8  &4161.9&  &\\
4  &[2,6,2]  &...                &        &4395.5&  &\\
5  &[3,6,1]  &$cs\bar{c}\bar{s}$ &        &4402.8&4383.4  &\\
6  &[3,6,2]  &$cs\bar{c}\bar{s}$ &        &4394.7&  &\\
\hline\hline
\end{tabular}
\end{center}
\end{table}

The $J^{PC}=0^{++}$ system: there are totally twelve channels. All results with $J^{PC}=0^{++}$ are given in Table \ref{0++}. From the table \ref{0++}, we observe that, under the meson-meson configuration, the energy of each individual color-singlet channel exceeds the corresponding theoretical threshold, indicating the absence of bound states for $c\bar{c}s\bar{s}$ system with $J^{PC}=0^{++}$. For the hidden-color channels, their energies are generally higher than those of the color-singlet channels. Under the diquark-antidiquark configuration, the energies of all individual channels exceed the lowest energy of the $D_s^+D_s^-$ system, with the lowest energy being 4305.3 MeV. Subsequently, we calculated the coupled channels under the meson-meson configuration, yielding an energy of 3788.5 MeV. Similarly, for the diquark-antidiquark configuration, the coupled channel energy was found to be 4249.0 MeV. We observed coupling between individual channels, but the coupling strength was not significant. Finally, we considered the coupling of all channels for both configurations, resulting in a coupling energy of 3788.4 MeV, which is almost identical to 3788.5 MeV. This indicates that the coupling strength between the two configurations is also very small.

The $J^{PC}=1^{++}$ system: there are 4 meson-meson channels and 2 diquark-antidiquark channels, totally 6 channels here. Table \ref{1++} lists the calculated masses of these channels and also their coupling results. From the table, we can draw conclusions similar to those in Table \ref{0++}. All individual channels are unbound. By coupling channels within the same configuration, the lowest mass for the meson-meson structure is 4115.3 MeV, and for the diquark-antidiquark structure, the lowest energy is 4326.4 MeV. Both two energies exceed the theoretical minimum threshold of $J/\psi\phi$, 4112.8 MeV. Furthermore, the lowest energy for the coupling of all channels is 4115.3 MeV, indicating that the coupling strength among all channels is relatively small here. And there is no any bound state in the $c\bar{c}s\bar{s}$ system with $J^{PC}=1^{++}$.

The $J^{PC}=1^{+-}$ system: there are 8 channels with a meson-meson structure and 4 channels with a diquark-antidiquark structure. Table \ref{1+-} presents the theoretical thresholds for each channel, with the lowest being the $J\psi\eta^{\prime}$ channel, which has a theoretical threshold of 3918.2 MeV. By calculating the lowest eigenenergy for each channel, we find that the energy is always higher than the corresponding threshold. When all meson-meson structure channels are coupled, the resulting energy is 3920.7 MeV, which is greater than 3918.2 MeV. When all diquark-antidiquark structure channels are coupled, the resulting energy is 4323.5 MeV, still exceeding the lowest threshold. If all 12 channels are coupled together, the lowest eigenenergy obtained is 3920.7 MeV. These results indicate that for the $1^{+-}$  state, no bound state is found.

The $J^{PC}=2^{++}$ system: Table \ref{2++} shows that there are two single-color channels ($J/\psi\phi$ and $D_s^{*+}D_s^{*-}$) and two hidden-color channels of the meson-meson structures, and two channels of the diquark-antidiquark structures for the $c\bar{c}s\bar{s}$ with $J^{PC}=2^{++}$ system. The situation is similar to the $J^{PC}=1^{++}$ system. The energy of each channel exceeds the corresponding theoretical threshold. At the same time, the channel coupling effect does not significantly lower the energy of this state. The lowest energy is still higher than the threshold of the lowest channel of $J/\psi\phi$, 4112.8 MeV. So, there is still no bound state for the $c\bar{c}s\bar{s}$ system with $J^{PC}=2^{++}$ in our present calculations. 

\begin{figure}
\center{\includegraphics[width=7.0cm]{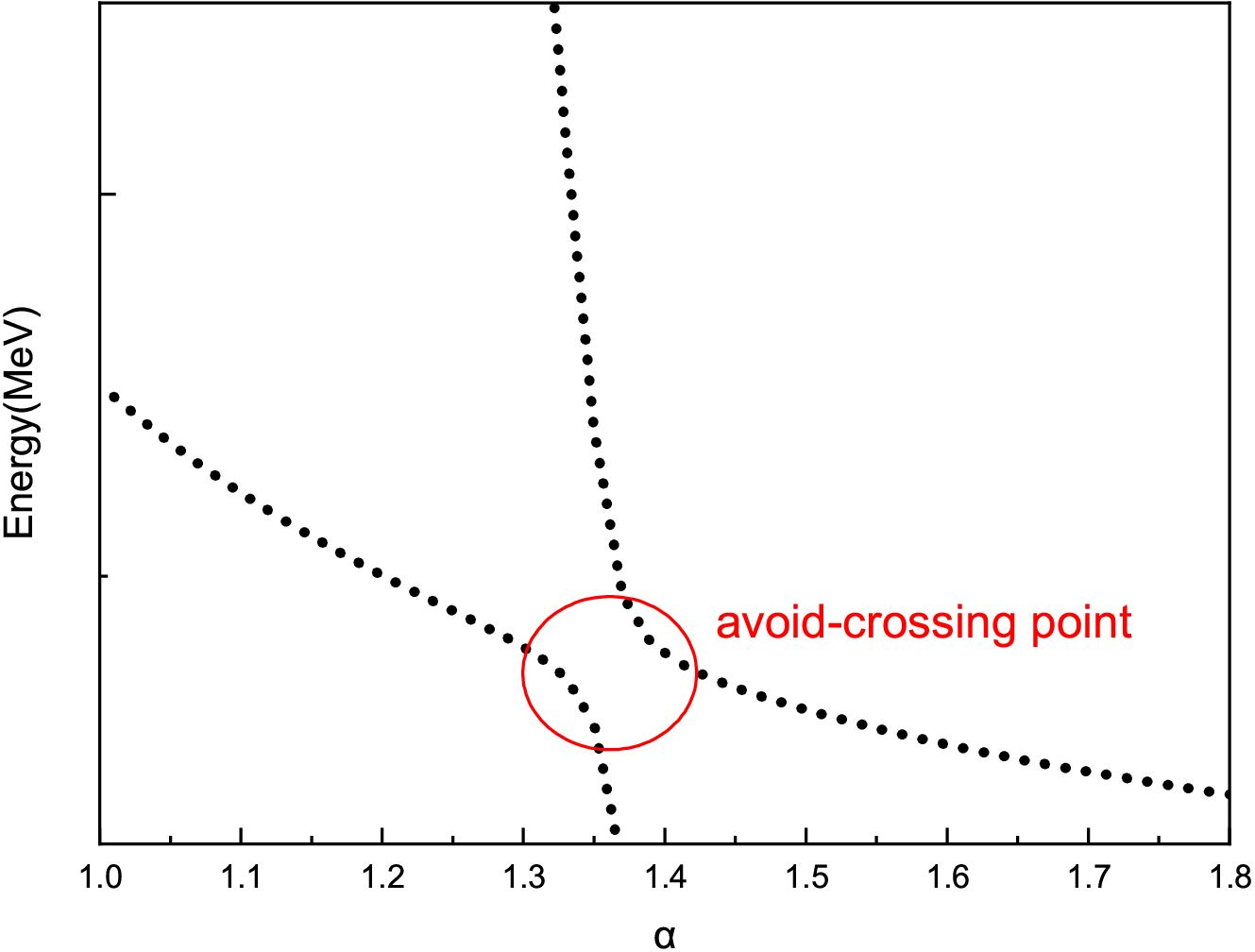}}
\caption{\label{avoid-crossing} Stabilization graph for the resonance.}
\end{figure}

\begin{figure}
\resizebox{0.50\textwidth}{!}{\includegraphics{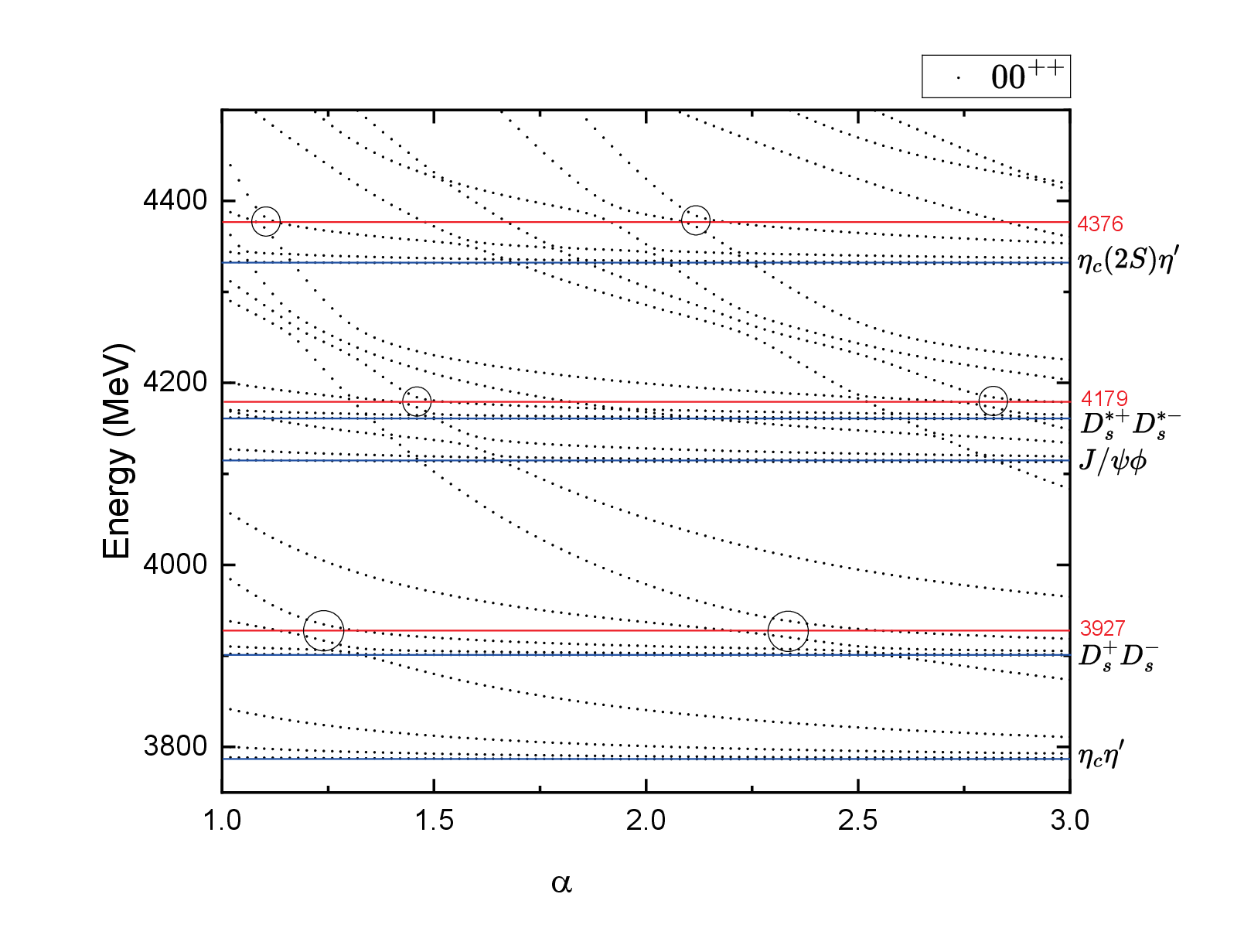}}
\caption{\label{rsm0++} The stabilization plots of the energies of $c\bar{c}s\bar{s}$ states for $J^{PC}=0^{++}$ with respect to the scaling factor $\alpha$.}
\end{figure}

\begin{figure}
\resizebox{0.50\textwidth}{!}{\includegraphics{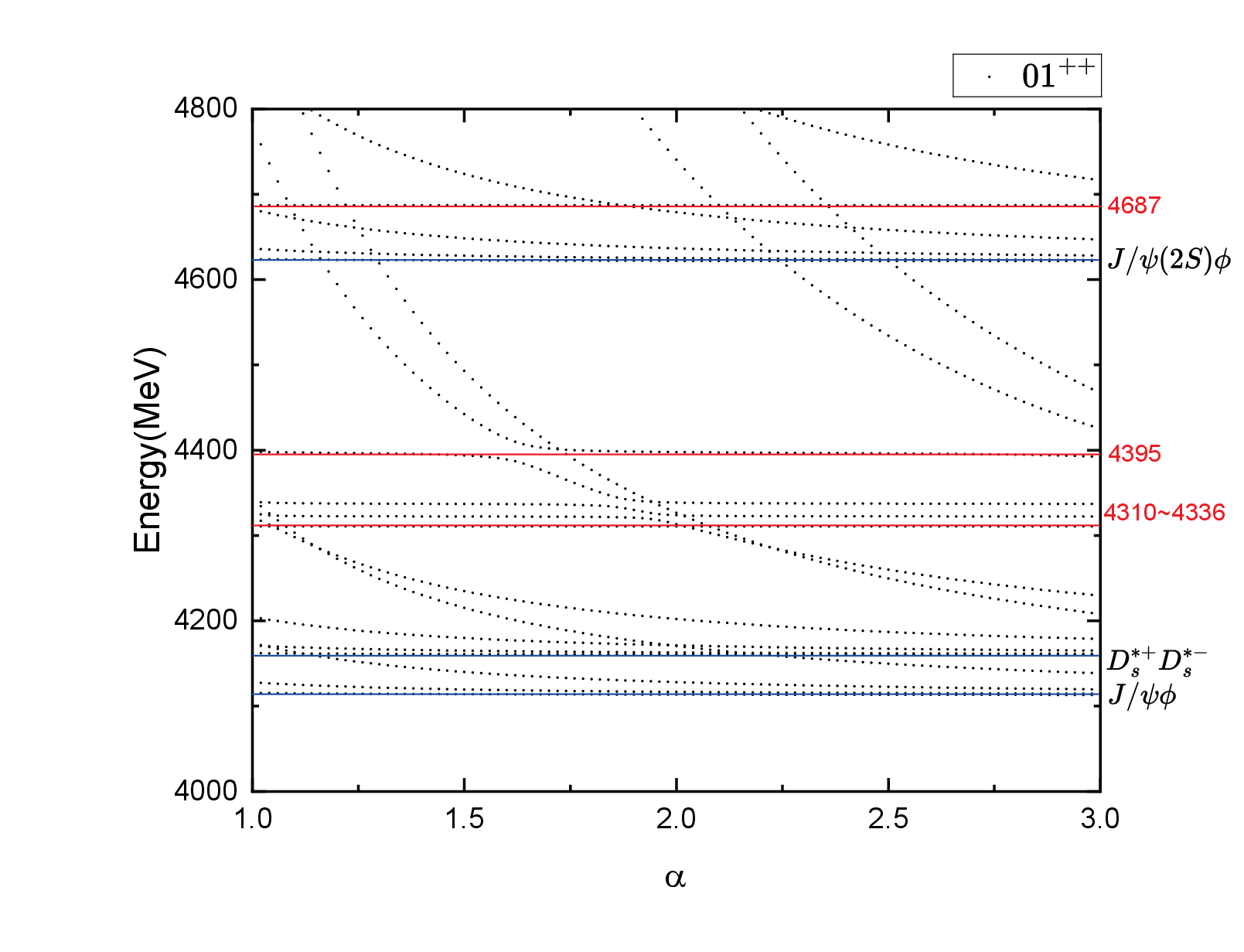}}
\caption{\label{rsm1++} The stabilization plots of the energies of $c\bar{c}s\bar{s}$ states for $J^{PC}=1^{++}$ with respect to the scaling factor $\alpha$.}
\end{figure}

\begin{figure}
\resizebox{0.50\textwidth}{!}{\includegraphics{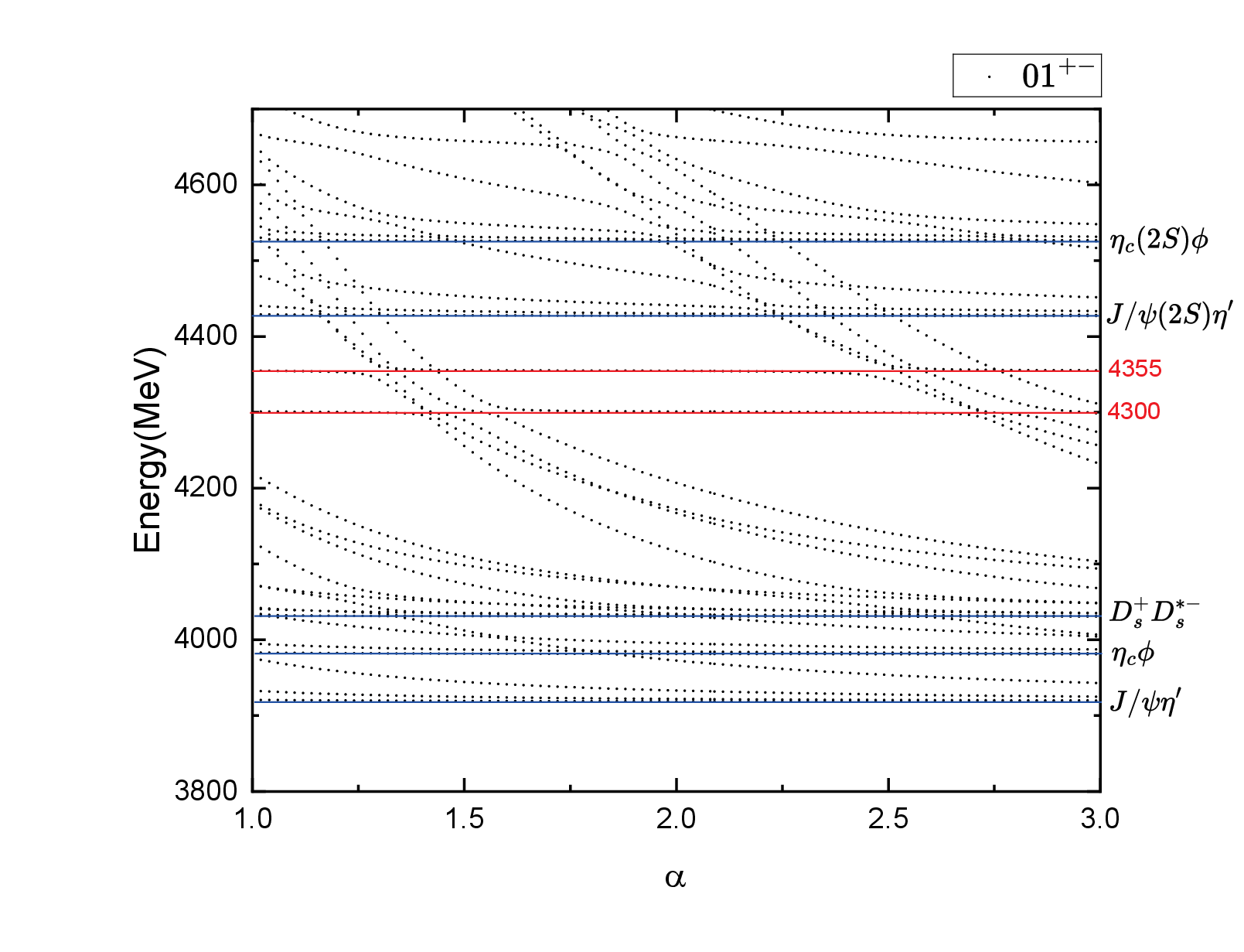}}
\caption{\label{rsm1+-} The stabilization plots of the energies of $c\bar{c}s\bar{s}$ states for $J^{PC}=1^{++}$ with respect to the scaling factor $\alpha$.}
\end{figure}

\begin{figure}
\resizebox{0.50\textwidth}{!}{\includegraphics{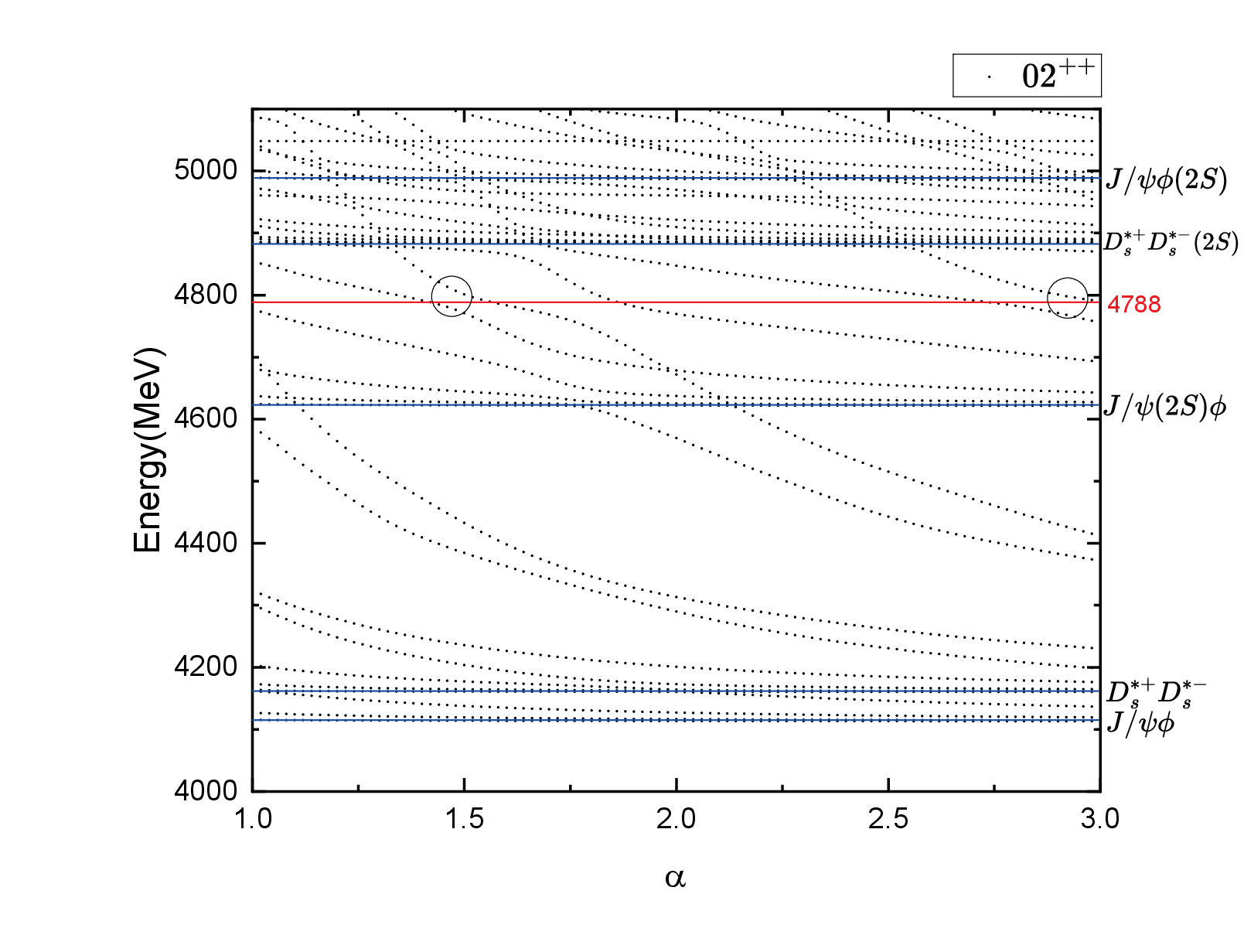}}
\caption{\label{rsm2++} The stabilization plots of the energies of $c\bar{c}s\bar{s}$ states for $J^{PC}=2^{++}$ with respect to the scaling factor $\alpha$.}
\end{figure}

Although no bound state is identified for the $J^{PC}=0^{++}$, $J^{PC}=1^{++}$, $J^{PC}=1^{+-}$ and $J^{PC}=2^{++}$ systems, the possibility of resonance states within the $c\bar{c}s\bar{s}$ tetraquark system cannot be excluded. Due to color confinement, the colorful subclusters: diquark and antidiquark, cannot separate directly, making the existence of resonance states feasible. To investigate the presence of such resonance states, we employ the stabilization method, a well-established technique for estimating the energies of stable states in electron-atom, electron-molecule, and atom-diatom complexes \cite{RSM}. In this approach, scaling the distance between two clusters causes the continuum states to converge towards their respective thresholds. Resonance states, on the other hand, exhibit stability if they do not couple to open channels, or manifest as avoided-crossing structures if they do couple, which is illustrated in Fig. \ref{avoid-crossing}. In the figure, the above line represents a scattering state, and it will fall down to the threshold. The line below is the resonance state, which tries to keep stable. The resonance state will interact with the scattering state, which will bring about an avoid-crossing point in the figure. Besides, these avoid-crossing structures reappear periodically with increasing scaling. This method has been successfully applied in studies of pentaquark systems \cite{Hiyama:2005cf, Hiyama:2018ukv}, and the tetraquark systems \cite{Chen:2019vrj, Chen:2021uou, Chen:2021crg, Jin:2020jfc}. To realize the real scaling method here, we multiply the Gaussian size parameter $r_n$ in Eq. \ref{gaussiansize} by a factor $\alpha$, $r_n \rightarrow \alpha r_n$ only for the meson-meson structure with color singlet-singlet configuration. Then we can locate the resonances of $c\bar{c}s\bar{s}$ system with respect to the scaling factor $\alpha$, which takes the values from 1.0 to 3.0. The results of the resonance state search for the $c\bar{c}s\bar{s}$ tetraquark state are demonstrated in Figs. \ref{rsm0++}, \ref{rsm1++}, \ref{rsm1+-} and \ref{rsm2++}, respectively.

For the $c\bar{c}s\bar{s}$ with $J^{PC}=0^{++}$ system, the results of the resonance search are presented in Fig. \ref{rsm0++}, focusing on the energy range from 3800 MeV to 4500 MeV. In the figure, five blue horizontal lines denote the theoretical thresholds of five channels, $\eta_c\eta^{\prime}$, $D_s^+D_s^-$, $J/\psi\phi$, $D_s^{*+}D_s^{*-}$, $\eta_c(2S)\eta^{\prime}$. Additionally, three red lines represent potential resonance states. Repeatedly occurring avoid-crossing points are marked with circles for clarity.
The first notable point appears above the $D_s^+D_s^-$ threshold, where a resonance state is identified at an energy of approximately 3927 MeV. This theoretical prediction is in excellent agreement with the mass of $X(3960)$ reported by the LHCb collaboration \cite{LHCb:2022aki}. As a result, our calculations suggest that $X(3960)$ can be interpreted as a $c\bar{c}s\bar{s}$ tetraquark state with quantum number $J^{PC}=0^{++}$.  
Another possible resonance state is found near the threshold above $D_s^{*+}D_s^{*-}$, with an energy of around 4179 MeV. This state likely corresponds to the $X(4140)$ observed by the LHCb experiment \cite{LHCb:2022aki}. Finally, a resonance state with an energy of approximately 4376 MeV is identified at a higher energy region. By comparing with the experimental results, we find that the energy of 4376 MeV is close to the $X(4350)$, and the quantum number $J^{PC}=0^{++}$ is consistent with the reported data by the Belle Collaboration \cite{Belle:2009rkh}.  
Our findings are consistent with those obtained using the Born-Oppenheimer approach, which predicted a mass of 4370 MeV \cite{Braaten:2014qka}. Additionally, Ref. \cite{Yang:2019dxd} identified $X(4350)$ as a strong candidate for a compact tetraquark state with $J^{PC}=0^{++}$ within the framework of the chiral quark model.

For the $J^{PC}=1^{++}$ system in Fig. \ref{rsm1++}, focusing on the energy range from 4000 MeV to 4800 MeV, we obtain three horizontal blue lines, each representing a threshold state. Additionally, around an energy of 4310 MeV, we observe three closely spaced horizontal lines, which predominantly correspond to hidden-color states with very weak coupling to color-singlet channels. Therefore, we suggest the existence of a hidden-color-dominated resonance in the 4310$\sim$4336 MeV region. Similarly, at approximately 4395 MeV and 4687 MeV, we identify two more horizontal resonance levels. These states are also primarily hidden-color states with minimal coupling to color-singlet channels; hence, their avoided crossings are not depicted in the figure.

For the $J^{PC}=1^{+-}$ system in Fig. \ref{rsm1+-}, in the energy range of 3800 MeV to 4700 MeV, we obtain five corresponding threshold lines, along with two resonance states at 4300 MeV and 4355 MeV. These two resonances also have very weak coupling to color-singlet channels and are predominantly hidden-color states.

For the lat system with $J^{PC}=2^{++}$, the resonances are shown in Fig. \ref{rsm2++}. The first two blue horizontal lines represent the theoretical thresholds of two channels, $J/\psi\phi$ and $D_s^{*+}D_s^{*-}$. Higher-energy channel thresholds are also indicated by blue lines in the figure, such as $J/\psi(2S)\phi$, $D_s^{*+}D_s^{*-}(2S)$, $J/\psi\phi(2S)$. In the present calculations, only one potential resonance state was identified for the $J^{PC}=2^{++}$ configuration, with an energy of approximately 4788 MeV. To date, no experimental evidence for this state has been reported, suggesting it could be a candidate for an exotic state.

\section{Summary}
\label{summary}
The $c\bar{c}s\bar{s}$ system with the quantum numbers $J^{PC}=0^{++}, 1^{++}, 1^{+-}, 2^{++}$ have been systemically studied in the framework of the chiral quark model(CQM). In the current work, our aim is to search for possible bound states or resonant states in this system and to explain the newly discovered exotic states observed in experiments. Two structures:  the meson-meson and diquark-antidiquark structures are taken into account. We also performed the single-channel and channel-coupling calculations. In order to search for any resonance state, a stabilization method - the real scaling method (RSM) is applied to the coupling calculation of all channels for both two configurations.

Our numerical results indicate that, whether using single-channel calculations or considering channel-coupling calculations, the system's lowest eigenvalue is greater than the corresponding theoretical threshold. No bound states were found for $J^{PC}=0^{++}, 1^{++}, 1^{+-}, 2^{++}$ system.
Additionally, regarding the search for resonant states, some meaningful results were obtained. For instance, in the $J^{PC}=0^{++}$ system, three possible resonant states were identified in the energy range of 3800 MeV to 4500 MeV, with energies of 3927 MeV, 4179 MeV, and 4376 MeV, respectively. Notably, the resonant state $X(3927)$ matches very well with the energy of the newly discovered exotic state $X(3960)$ reported by the LHCb collaboration \cite{LHCb:2022aki}. As a result, our calculations suggest that $X(3960)$ can be interpreted as a $c\bar{c}s\bar{s}$ tetraquark state with quantum number $J^{PC}=0^{++}$. Another possible resonance state with 4179 MeV likely corresponds to the $X(4140)$ observed by the LHCb experiment \cite{LHCb:2022aki}. And the resonance with 4376 MeV is close to the $X(4350)$ reported by Belle Collaboration \cite{Belle:2009rkh}.
For the $J^{PC}=1^{++}$ system in the energy range from 4000 MeV to 4800 MeV, there is likely one resonance in the energy range of 4310$\sim$4336 MeV, along with two additional resonances at 4395 MeV and 4687 MeV, respectively.
And for the $J^{PC}=1^{+-}$ system, there exist two resonance states in the 4200$\sim$4400 MeV range in our present work, with energies of 4300 MeV and 4355 MeV, respectively.
For the $J^{PC}=2^{++}$ system, a resonant state was identified in the energy range from 4000 MeV to 5100 MeV, with an energy of approximately 4788 MeV.

All of these resonant states merit further experimental investigation. We recommend conducting additional experimental tests to verify the existence of these potential resonance states. Furthermore, to confirm the presence of these $c\bar{c}s\bar{s}$ tetraquarks, future studies should focus on investigating the scattering processes of the corresponding open channels.

\section{Acknowledgment}
This work is partly supported by the National Natural Science Foundation of China under Grants No. 12205125, No. 11847145, No. 12205249 and No. 11865019, and also supported by the Natural Science Foundation of Jiangsu Province under Grants No. BK20221166.

\section{ Data Availability Statement}
Data Availability Statement This manuscript has no associated data or the data will not be deposited. [Authors' comment: The data have been illustrated in the figures and tables, so they are not necessary to be deposited. Data may be made available upon request.]

\end{document}